\documentclass[10pt,conference]{IEEEtran}
\IEEEoverridecommandlockouts

\usepackage{cite}
\usepackage{float}
\usepackage{amsmath,amssymb,amsfonts}
\usepackage{algorithmic}
\usepackage{graphicx}
\usepackage{textcomp}
\usepackage{xcolor}
\usepackage{hyperref}
\usepackage{booktabs}
\usepackage{multirow}

\newcommand{\partner}{Nullify}
\newcommand{\repojava}{Spring Support Portal}

\usepackage[most]{tcolorbox}

\newtcolorbox{rqanswer}[1]{
  enhanced, breakable=false, arc=2pt, boxrule=0.6pt,
  colback=black!4, colframe=black!75,
  left=6pt, right=6pt, top=5pt, bottom=5pt,
  coltitle=white, colbacktitle=black!75,
  fonttitle=\bfseries\footnotesize, title={#1}
}

\def\BibTeX{{\rm B\kern-.05em{\sc i\kern-.025em b}\kern-.08em
    T\kern-.1667em\lower.7ex\hbox{E}\kern-.125emX}}
\begin{document}

\title{GraftyVul: Synthesising Insecure Programs Through Real-World Vulnerability Grafting}

\author{
    \IEEEauthorblockN{
        Omri Ram\IEEEauthorrefmark{1}\IEEEauthorrefmark{2},
        Mitchell Horner\IEEEauthorrefmark{1}\IEEEauthorrefmark{2},
        Ron Van der Meyden\IEEEauthorrefmark{1},
        Alsharif Abuadbba\IEEEauthorrefmark{3},
        Hammond Pearce\IEEEauthorrefmark{1}
    }
    \IEEEauthorblockA{\IEEEauthorrefmark{1} University of New South Wales, Australia}
    \IEEEauthorblockA{\IEEEauthorrefmark{2} Nullify, Australia}
    \IEEEauthorblockA{\IEEEauthorrefmark{3} CSIRO, Australia}
    \IEEEauthorblockA{
        \{o.ram, mitch.horner, r.vandermeyden, hammond.pearce\}@unsw.edu.au,
        sharif.abuadbba@csiro.au
    }
}

\maketitle

\begin{abstract}
Vulnerability datasets underpin a wide range of security research, including vulnerability detection, automated remediation, and secure code generation. However, existing datasets sacrifice at least one of three desirable properties: diversity (of language or vulnerability type), reproducibility/executability, or realism. We therefore present GraftyVul, a system that constructs vulnerable programs by grafting real-world vulnerabilities into open-source projects. This grounds the dataset in vulnerabilities observed in real-world contexts while harnessing known good build and test environments, enabling exploit-verification scripts to guarantee that an introduced vulnerability successfully alters a program’s behaviour. Using GraftyVul, we generate 212 verified and exploitable vulnerable programs spanning five programming languages (Python, TypeScript, Java, Go, and
C\#) across 23 CWE categories. To evaluate fidelity, we introduce a language- and context-agnostic semantic embedding that compares vulnerabilities by sink, mechanism and host-feature rather than surface code. This approach outperforms standard code embeddings on cross-language clone and CWE classification. These embeddings demonstrate that GraftyVul samples retain a strong semantic signature to their source vulnerability. We additionally compare GraftyVul against 13 widely used datasets, where it attains competitive diversity while being the only reproducible-exploit dataset with broad language and CWE coverage. Finally, we illustrate GraftyVul’s practical utility through an industrial case study evaluating a production vulnerability remediation system.
\end{abstract}


\section{Introduction}
The growing complexity and ubiquity of software in daily life amplifies the risks associated with cybersecurity breaches \cite{cobos2024review}. Much of cybersecurity research depends on high quality datasets for benchmarking, building models, and improving detection and remediation tools~\cite{yinThinkRepairSelfDirectedAutomated2024a, le-congReliableEvaluationNeural2025b, jimenezImportanceAccountingRealworld2019}. The breadth and quality of these datasets limits the research they enable. Vulnerability datasets often support few languages~\cite{wangPatchDBLargeScaleSecurity2021, buiVul4JDatasetReproducible2022b}, do not contain a mechanism to trigger the vulnerability, known as a proof-of-concept (PoC) exploit \cite{bhandariCVEfixesAutomatedCollection2021, quan2025empirical, nikitopoulosCrossVulCrosslanguageVulnerability2021a},  or are limited to a narrow range of vulnerabilities \cite{meiARVOAtlasReproducible2024, dolan-gavittLAVALargeScaleAutomated2016}. 

Synthesising vulnerabilities provides an avenue for curating high quality reproducible benchmarks. Existing methods for synthesising vulnerability data fall into three main categories. \emph{Mutation} approaches transform secure code by applying known vulnerability patterns~\cite{nongVGXLargeScaleSample2024b, nongVULGENRealisticVulnerability2023, nie2026secodeplt}. \emph{Injection} approaches insert vulnerabilities into existing programs~\cite{dolan-gavittLAVALargeScaleAutomated2016, pewnyEvilCoderAutomatedBug2016}, while \emph{pure synthesis} approaches generate vulnerable code entirely from scratch~\cite{hajipourHexaCoderSecureCode2024, boland2012juliet}. These approaches span mutation, injection, and pure synthesis, and divide methodologically into rule-based and learning-based techniques. More recently, LLM-based methods have leveraged the reasoning capabilities of large language models to synthesise vulnerability data~\cite{ullah2025cveentriesverifiableexploits, hajipourHexaCoderSecureCode2024, daneshvar2026vulscriber}, using agentic workflows that decompose complex tasks and iteratively refine outputs through self-verification. Despite their differences, many of these approaches all create vulnerabilities in artificial, templated, or otherwise synthetic code, rather than in the real software where vulnerabilities naturally arise. Additionally, they might not have sufficiently robust vulnerability verification tests -- giving rise to a lower quality dataset. Consequently, the resulting samples may lack the verifiability and complexity of vulnerabilities encountered in practice. 

In this work, we are likewise interested in creating synthetic vulnerabilities using an agentic paradigm. However, rather than generating vulnerabilities from patterns, templates, or from scratch, our approach \textit{grafts} real-world vulnerabilities sourced from an industry partner (\partner{}) into pre-sourced and curated open-source projects. \partner{},  a commercial security vendor, obtains these vulnerabilities by scanning its clients' software. 

We employ pre-existing agentic infrastructure to characterise the identified vulnerabilities before feeding them into GraftyVul. This system grafts the vulnerability into a functional target program rather than synthesising it in isolation. By using high-quality target environments which have pre-existing tests and verifiers, GraftyVul can ensure that original target functionality remains intact and that build scripts can still compile and execute the program after the vulnerability is inserted. Then, our exploit verification script ensures that the vulnerability is inserted correctly by providing a ``PoC'' which can retrieve or modify a seeded secret through the injected vulnerability.

To measure the relationship between the newly synthesised vulnerability sample and the original real-world vulnerability, we can adopt code clone detection methods that demonstrate the semantic fidelity between the grafted and source vulnerabilities. However, existing methods for semantic code clones do not provide a sufficiently broad definition for detecting vulnerable code clones across contexts. Vulnerable code clone tools may employ graphs to characterise the vulnerability, which rely heavily on syntax and similar business logic, which does not translate well into cross-context scenarios ~\cite{feng2024fire}. Additional methods employ embeddings~\cite{shimmi2024vulsim} which do not necessarily capture the fine grained characteristics our vulnerability grafts exhibit. Thus, we produce alternative semantic embeddings for demonstrating the semantic signature preserved in the synthesised samples when compared to the original vulnerability. We employ this metric to determine the intra-vulnerability similarity to determine if we are creating a diverse dataset; and likewise can compare this diversity against other existing benchmarks. Finally, we demonstrate GraftyVul's utility in evaluating \partner{}'s deployed LLM-driven autofix tool employed for vulnerability remediation. 

In summary, we provide the following contributions:
\begin{itemize}
    \item \textbf{GraftyVul}, an LLM-driven system that synthesises vulnerabilities analogous to real-world industry findings, each accompanied by a verified exploit. \textit{[GraftyVul will be made open-source upon publication.]}
    \item \textbf{A semantic similarity embedding approach} that characterises a vulnerability sample in a language- and context-agnostic way.
    \item \textbf{A comparison of GraftyVul} against 13 other widely used vulnerability datasets across the properties relevant to security research and their diversity.
    \item \textbf{A case study} demonstrating the utility of GraftyVul's synthesised samples in production.
\end{itemize} 

We find that GraftyVul maintains a strong semantic signature to the original industry vulnerability using the developed semantic similarity embedding, achieving \textbf{16\% recall@1} against a 0.47\% random baseline, while human evaluators can map the synthesised vulnerability to its source finding with an \textbf{accuracy of 59.5\%}, approximately six times chance. Additionally, we utilize the Vendi score~\cite{friedman2023vendi} on GraftyVul's semantic embedding to derive a score of 102 compared to a median of 109.2, more diverse than all but one reproducible dataset while being the only reproducible dataset with broad Common Weakness Enumeration (CWE) and language coverage across 13 other widely used vulnerability datasets, additionally providing executable tests and a verified PoC for each sample. Through benchmarking \partner{}'s deployed autofix tool on GraftyVul samples we identify recurring and actionable failure modes, including fixes that are secure but break legitimate functionality, and fixes that close the reported weakness but leave the underlying exposure intact.

\section{Background and Related Work}

\subsection{Vulnerability Datasets and Benchmarks}
\subsubsection{Mined Datasets}
The National Vulnerability Database (NVD) catalogues known common vulnerabilities and exposures (CVE) together with metadata such as affected versions and patch availability~\cite{booth2015national}. Mined datasets assemble vulnerable code and its fixes from the NVD and related sources. CVEfixes~\cite{bhandariCVEfixesAutomatedCollection2021} collects CVEs from the NVD and extracts the corresponding fix commits and changed functions from the referenced repositories, while PatchDB~\cite{wangPatchDBLargeScaleSecurity2021} assembles a large set of security patches drawn both from patches referenced in the NVD and from GitHub commits identified through a nearest-link search. DiverseVul~\cite{chen2023diversevul} crawls security issue websites to identify vulnerability-fixing commits and their functions. Mining real-world sources in this way yields broad coverage across languages and CWEs. However, these datasets are label-only, and because their quality is bounded by the mined sources, many contain noisy, irrelevant code and untrustworthy vulnerability labels~\cite{chen2023diversevul, ding2024vulnerability}.

\subsubsection{Curated Datasets}
Curated datasets therefore refine the mined datasets to improve label quality and derive additional metadata. PyVul~\cite{quan2025empirical} targets the Python package ecosystem, providing publicly reported, developer-verified vulnerabilities linked to their affected packages, each paired with a vulnerable and a patched Python function, with LLM-assisted cleaning applied to raise function-level label accuracy to 94\%. CleanVul~\cite{liCleanVulAutomaticFunctionLevel2025} applies LLM-based heuristics to separate genuine vulnerability-fixing changes from the unrelated edits bundled into the same commit, reducing the label noise that affects mined datasets. PrimeVul~\cite{ding2024vulnerability} consolidates and de-duplicates samples drawn from several earlier datasets and applies a stricter labelling pipeline to raise label accuracy and remove duplication. While these efforts improve the reliability of the labels, the resulting datasets remain label-only, still offering no means to build the affected software or confirm that a vulnerability is exploitable.

\subsubsection{Reproducible Datasets}
Reproducible datasets may be curated in such a way that allows the vulnerability to be exercised. VulnRepairEval's~\cite{wang2025vulnrepaireval} authors hand-picked CVEs and produced a PoC for each, using differential testing to evaluate LLM-based AVR tools. Vul4J~\cite{buiVul4JDatasetReproducible2022b} mines real Java vulnerabilities, keeping only those that are reproducible, each providing a proof-of-vulnerability test that fails on the vulnerable version and passes on the fixed one. VJBench extends this with 42 further reproducible Java vulnerabilities mined from the NVD that Vul4J does not cover~\cite{wuHowEffectiveAre2023}. ARVO~\cite{meiARVOAtlasReproducible2024} mines OSS-Fuzz~\cite{Arya_OSS-Fuzz}, a Google service that fuzzes open-source projects, identifying the commits that introduce and fix each bug. It provides an interface to rebuild the affected C/C++ project at either version together with the input that triggers the bug, giving both a build mechanism and a proof of exploit. 
These datasets achieve reproducibility but each covers a single language and, in ARVO's case, only identifies and reproduces those memory-safety bugs surfaced by fuzzing.

\subsection{Synthetic Vulnerability Benchmark Generation}

\subsubsection{Rule-based and Template Generation}
Rule-based methods construct vulnerable samples from hand-written patterns or by transforming existing code. The Juliet test suite~\cite{boland2012juliet} synthesises template C and Java programs from pre-defined vulnerability patterns, each built around a specific control- or data-flow that carries the flaw. SARD~\cite{black2018software}, the reference dataset that hosts Juliet and similar suites, builds synthetic cases by taking known vulnerabilities and generating additional code around them to increase complexity. LAVA~\cite{dolan-gavittLAVALargeScaleAutomated2016} injects bugs into open-source C programs by identifying points on an execution trace where input bytes are available but do not influence control flow, then inserting a bug that fires only when those bytes take a specific value. EvilCoder~\cite{pewnyEvilCoderAutomatedBug2016} injects vulnerabilities into open-source repositories by deriving a graph representation of the code to trace source-to-sink paths, then removing sanitisation checks and other guarding logic that would otherwise prevent exploitation. While allowing for synthesis of enormous datasets, these datasets tend to lack diversity and realism encountered in real-world scenarios.

\subsubsection{Learning-based Methods}
Learning-based methods use neural models, and increasingly large language models, to generate, inject, or reproduce vulnerable code. CVE-Genie~\cite{ullah2025cveentriesverifiableexploits} reproduces real CVEs through an agentic workflow that reconstructs the build environment and synthesises a PoC exploit. HexaCoder~\cite{hajipourHexaCoderSecureCode2024} generates vulnerable and secure code pairs from security reports and validates the vulnerability through a static analyser. SeCodePLT~\cite{nie2026secodeplt} applies LLM-guided mutations to seed CWE examples to expand a dataset of vulnerabilities. VulGen~\cite{nongVULGENRealisticVulnerability2023} uses a neural model to identify vulnerability injection sites in source code and applies code-transformation patterns mined from historical fixes to inject vulnerabilities. VGX~\cite{nongVGXLargeScaleSample2024b} similarly learns where to inject the vulnerabilities but draws its transformation patterns from both historical fixes and human knowledge of real-world vulnerabilities. Across these methods verification ranges from none, where the output is label-only as in VulGen and VGX, through static analyser checks in HexaCoder and executable tests in SeCodePLT, to model judgement in CVE-Genie.

\subsection{Code Similarity and Clone Detection}
Code clones are conventionally graded from Type-1, exact copies, through Type-2 and Type-3, renamed and modified copies, to Type-4 clones, which are semantically equivalent but syntactically unrelated~\cite{roy2007survey}. Grafting a vulnerability from one context to another seeks to produce Type-4 vulnerable code clones. Structural methods address the easier cases efficiently. VUDDY~\cite{kimVUDDYScalableApproach2017} abstracts each function and matches a hash of its normalised body, scaling to large codebases but remaining sensitive to syntactic, language, and contextual variation. FIRE~\cite{feng2024fire} performs taint-flow analysis to track taint propagation and represents each function as a vector over its semantic, lexical, and syntactic features. A second family learns code representations directly, embedding functions with models such as VulSim~\cite{shimmi2024vulsim} to retrieve similar vulnerabilities by vector distance. Because all of these methods derive their representation from code structure or learned code embeddings, their similarity is dominated by the syntax and language of the samples rather than the underlying vulnerability, and degrades once language or context varies. Sannini et al.~\cite{sannini2026identifying} instead detect vulnerable code clones from LLM-generated summaries, showing that natural-language descriptions can capture context-agnostic vulnerability semantics, though they evaluate only on Solidity contracts. No existing method is suited to measuring the similarity between a source and grafted vulnerability, which motivates the semantic embedding approach we introduce in Section~\ref{subsec:sem-sim}.

\section{GraftyVul System Design}
\begin{figure}[tbp]
  \centering
  \includegraphics[width=0.8\columnwidth]{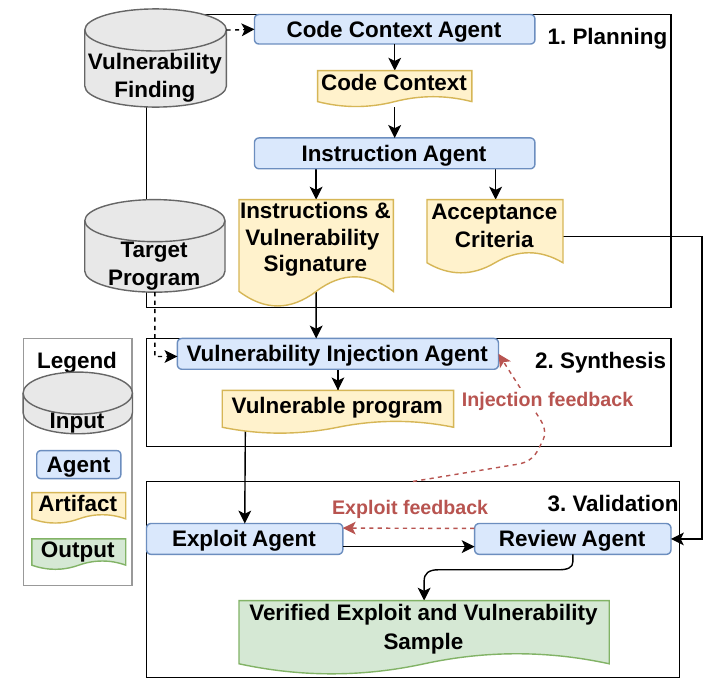}
  \caption{GraftyVul overview. The five LLM-driven agents are grouped into three modules: planning, synthesis, and validation, with a feedback loop from validation back to synthesis.}
  \label{fig:overview}
\end{figure}
\subsection{Overview}
GraftyVul provides a new method for synthesising a vulnerability dataset, reproducing real-world vulnerability findings by grafting them into known-good target programs. Given a vulnerability finding and a target program, GraftyVul coordinates five LLM-driven agents to synthesise, exploit, and verify a semantically analogous vulnerability in the target. Each agent is a large language model given a specific role, a set of tools it can invoke, and the autonomy to decide how to apply them. GraftyVul's workflow is illustrated in Figure~\ref{fig:overview}.

The five agents form three modules: planning, synthesis, and validation, with a feedback loop between validation and synthesis. The code context agent derives the data flow and mechanism of the source vulnerability; the instruction generation agent translates this into acceptance criteria and a vulnerability signature; the vulnerability injection agent synthesises the candidate vulnerable code; the exploit agent produces a proof-of-exploit; and the review agent verifies the fidelity of the graft to the source vulnerability. Failed verification triggers iteration, and the pipeline terminates either on success, after three rejections from the validation module, or on surrender of the instruction or exploit agents. The following sections provide an overview of the target programs, role of each agent in GraftyVul and the deterministic guard rails ensuring that synthesised vulnerabilities are exploitable.

\subsection{Target Programs}
Rather than synthesising vulnerable code from scratch, GraftyVul grafts known vulnerabilities onto pre-existing functional repositories. This provides two benefits: a tailored implementation of the surrounding feature that mimics the original vulnerability finding, and a pre-existing environment for regression testing and exploit verification.

GraftyVul ensures the functionality and exploitability of synthesised samples by integrating a target program via three pre-curated scripts. A setup script compiles and prepares the target application, seeding it with at least one flag string placed in a location inaccessible through normal application use. An exploit validation script verifies that a given flag has been retrieved or altered through an exploit agent's PoC, indicating a successful exploit. A test script invokes the repository's regression suite to confirm that pre-existing functionality is preserved after the vulnerability patch is applied. %

\subsection{Planning}
This section outlines the five subagents comprising the three submodules employed in GraftyVul, as shown in Figure~\ref{fig:overview}. %

\paragraph{Code Context Agent}
Integrated in \partner{}'s platform is a code context agent which characterises identified vulnerabilities. We provide the code context agent's output into GraftyVul; though we note any similar input could be fed into our system given the following characteristics.
Starting from a Static Application Security Testing (SAST) finding (a vulnerability detected by static analysis tooling, with sink location and CWE metadata), the code context agent clones the source repository and traces the data flow from potential sources to the reported sink using language server tooling. The sink is the program location where untrusted data reaches the sensitive operation that realises the weakness and the source is the point where that untrusted data first enters the program, and the data flow connecting them is the path the vulnerability exploits. The agent collects the code blocks along this trace and also provides a natural-language description of the vulnerability mechanism and the underlying feature. The resulting bundle, comprising the code blocks, mechanism description, and SAST metadata, is hereafter referred to as the `code context' of the finding.

\paragraph{Instruction Generation Agent}
Given a target repository and the code context, the instruction agent devises a plan to introduce a vulnerable feature analogous to the source vulnerability. It employs code exploration tools to collect information pertaining to the target program's structure and functionality, then produces three artefacts consumed at downstream stages. First, step-by-step instructions for extending the repository to introduce the specified vulnerability in an analogous feature. Second, a vulnerability signature capturing the mechanism, sink, and data-flow of the source vulnerability. Third, acceptance criteria provided to the review agent and withheld from the injection agent. The instruction agent may also terminate the workflow if it determines that the vulnerability cannot be faithfully implemented in the target program.

\subsection{Synthesis}
\paragraph{Vulnerability Injection Agent}
The vulnerability injection agent implements the vulnerable feature specified by the instruction agent. It uses language server tools to navigate and edit the target program and may execute the repository's test suite to verify that the project still builds and that pre-existing functionality is preserved alongside the introduced vulnerability. The agent may submit its vulnerable patch only after all tests pass. This constraint ensures that the host repository's non-vulnerable functionality is not regressed. Upon a failed review or exploit, the agent receives feedback specifying which criteria were not met and iterates accordingly up to three times. 

\subsection{Validation}
\paragraph{Exploit agent}
The exploit agent produces a proof-of-concept exploit for the vulnerability synthesised by the injection agent. To ensure that the exploit genuinely triggers the introduced vulnerability, the target program's setup script seeds a hidden flag (a secret value injected at build time, inaccessible through the application's normal interface). A valid exploit must retrieve or modify this flag through the vulnerable feature synthesised by the injection agent. The exploit agent may submit its script only after it is verified to fail against the secure repository and succeed against the repository with the vulnerable patch. The target program thus must have an exploit verification script which checks if the hidden flag has either been altered or successfully exfiltrated. This prevents the agent from finding workarounds not utilising the exploit. When no working exploit can be produced, the agent provides feedback to the injection agent and returns to the synthesis stage of GraftyVul.

\paragraph{Reviewer Agent}
The review agent receives the vulnerability patch, the exploit script, the acceptance criteria, the code context, and the vulnerability signature. It judges the produced sample against three criteria. First, the exploit script must not bypass the vulnerability mechanism and genuinely employ the introduced vulnerability to retrieve or alter the flag. Second, the sample must satisfy all acceptance criteria specified by the instruction agent. Third, the produced vulnerability must be analogous to the source code context, matching the mechanism, sink (as outlined by the vulnerability signature), and feature role. If any criterion fails, the review agent emits feedback identifying the violated criteria, which is routed to the injection or exploit agent as appropriate for re-synthesis.

\subsection{Implementation}
\label{sec: implementation}
GraftyVul is deployed on AWS, with agents invoking LLMs through Anthropic's native API. The system runs on Linux amd64 containers orchestrated via Fargate Step Functions. Target programs that require a database backend are deployed alongside a PostgreSQL sidecar. Agent orchestration is implemented with LlamaIndex's event-driven framework. Code exploration and editing across target programs are mediated by a custom Go-based Language Server Protocol layer that supports multiple LSP backends, exposed to agents via a unified gRPC interface providing find references, go to definition, and symbol lookup operations.

All agents operate at temperature 0 except the instruction agent, whose temperature is varied across ablations. We use Claude Sonnet 4.6 for the code context, exploit, and review agents. The instruction and injection agents are both ablated across Sonnet and Haiku.

\section{Evaluation}
In this section we evaluate GraftyVul alongside a novel semantic embedding used to measure similarity between vulnerability samples. We ablate relevant configurable components and measure semantic faithfulness to the source finding. We evaluate the anonymisation of the final dataset prior to publication to protect leakage of information of \partner{}'s clients. We compare GraftyVul's dataset with other widely used vulnerability datasets in terms of diversity and other characteristics pertinent to cybersecurity research. Finally, we demonstrate GraftyVul's usage at \partner{}. Our evaluation is framed around the following research questions:

\begin{description}
    \item[RQ1] Do LLM summaries faithfully capture vulnerability semantics and can these summaries be used to measure cross-context vulnerability similarity?
    \item[RQ2] Can GraftyVul synthesise and verify a vulnerability dataset semantically analogous to industry findings?
    \item[RQ3] How does GraftyVul's synthesised dataset compare to other available vulnerability datasets in regards to diversity and verifiability?
    \item[RQ4] Can GraftyVul's synthesised samples be utilised for benchmarking Automated Vulnerability Repair tools?
\end{description}

Table \ref{tab:target-programs} outlines the target programs used in this study to host the grafted vulnerabilities. We source three open source repositories spanning several languages and frameworks, as well as produce and host two additional programs to expand coverage. \textit{[Note that the two curated repositories will be released upon paper acceptance]}

\begin{table}[tbp]
\caption{Target programs used for hosting grafted vulnerabilities sourced from industry findings.}
\label{tab:target-programs}
\centering
\resizebox{\columnwidth}{!}{
\begin{tabular}{@{}p{1.5cm}llp{1.4cm}l@{}}
\toprule
\textbf{Repository} & \textbf{Language} & \textbf{Domain} & \textbf{License} & \textbf{Commit} \\
\midrule
\href{https://github.com/adr1enbe4udou1n/fastapi-realworld-example-app.git}{fastapi} & Python & Web API & MIT &\texttt{df1edfe} \\
\href{https://github.com/lattexi/express-ts-sqlite.git}{express-ts} & TypeScript & Media publishing & Apache-2.0 & \texttt{ec16111} \\
\href{https://github.com/gbrayhan/microservices-go.git}{microservices} & Go & Monolithic API & MIT & \texttt{b63ba3d} \\
\textit{spring}  & Java & REST API & MIT  & \texttt{N/A} \\
\textit{eshop}  & C\# & eCommerce   & MIT  & \texttt{N/A} \\
\bottomrule
\end{tabular}}
\end{table}

Through \partner{} we derive a collection of vulnerability findings detected by SAST tools. Integrated in \partner{}'s platform is the code context agent which characterises the vulnerability. The default GraftyVul agent configuration is set to use Sonnet 4.6 with a temperature of 0. We ablate these parameters to understand the impact of temperature and model selection on output. Following ablations, we execute GraftyVul to construct an evaluation dataset. Next, this dataset is compared against other widely used vulnerability datasets and benchmarks \partner{}'s automated vulnerability remediation tool. 

\subsection{Generation}
\label{subsec:generation}
We assess GraftyVul's capacity to synthesise verified vulnerability samples and characterise where synthesis completes. We define a sample to be verified when it adheres to the acceptance criteria specified by the instruction agent and the exploit verification scripts confirm the exploit agent's PoC. From the final dataset, we record for each attempt whether the pipeline produced a sample, the cost, and the status of any grafting attempts which are deemed infeasible. We report the synthesis rate and termination mode by CWE class and target language to identify which vulnerability types and languages the framework reproduces reliably. Furthermore, the fidelity of each verified sample to its source finding is quantified using the semantic similarity metric (Section~\ref{subsec:sem-sim}).

\begin{figure}[t]
  \centering
  \includegraphics[width=\columnwidth]{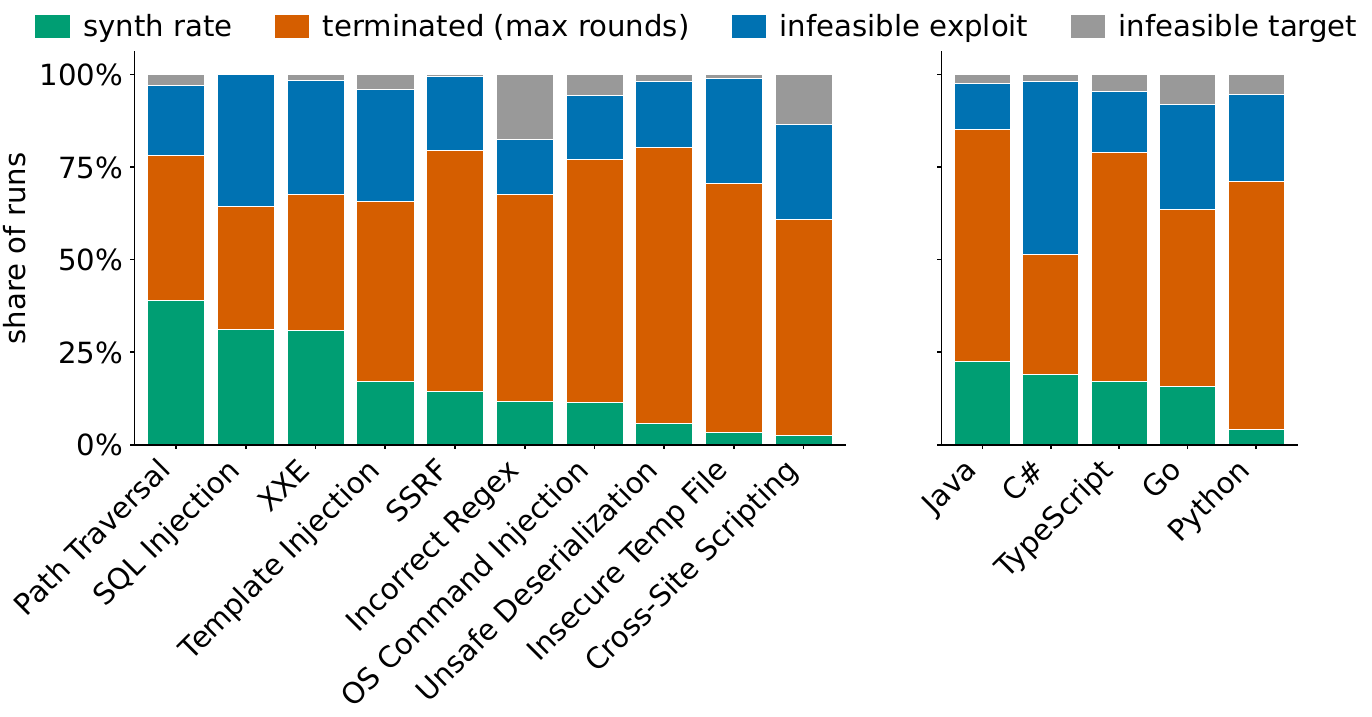}
  \caption{Synthesis rate and termination mode of the GraftyVul pipeline per CWE (the 10 most prevalent of 35 attempted) and language. Terminated and infeasible outcomes are cases where the validation module prevented a low-quality sample from being synthesised.}
  \label{fig:failure-mode-breakdown}
\end{figure}

Of the 35 distinct CWEs attempted, 23 were successfully produced. Of the 1361 executions that completed without infrastructure failure (a further 197 timed out from API rate limiting and were excluded), 301 (22.1\%) produced a vulnerability sample and PoC. Human inspection of these 301 found that 89 (29.6\%) exploited a weakness in the oracle rather than the target program. Although the flag is seeded at build time and withheld from the agents, in these cases the injection agent located the flag in the running environment and hard-coded it into the vulnerable feature itself, so the exploit agent could retrieve it directly. The sample then passes exploit verification without a genuine vulnerability being present, because the flag is exposed by the planted code rather than by a real weakness in the host program. These contrived CTF-style samples satisfy the flag-retrieval check while failing to represent an authentic vulnerability, so we exclude them from the dataset. This leaves a final dataset of 212 samples which do not hard-code the flag into the vulnerability. Figure~\ref{fig:failure-mode-breakdown} shows the termination modes which did not produce vulnerabilities across languages and CWEs. Overall execution cost had a mean of \$2.69 with a standard deviation of \$1.21, with failed runs costing a mean of \$3.16 and successful runs \$1.98. We evaluate the semantic similarity of the source vulnerability finding to the synthesised sample under section~\ref{subsubsec:retrieval-evaluation}.

\subsection{Validation}
\begin{figure*}[tbp]
\centering
\lstset{
  basicstyle=\footnotesize\ttfamily,
  breaklines=true, columns=fullflexible,
  keepspaces=true, showstringspaces=false,
  keywordstyle=\color{violet},
  stringstyle=\color{orange!80!black},
  commentstyle=\itshape\color{gray},
  aboveskip=2pt, belowskip=2pt,
}
\newcommand{\panelhd}[1]{\textbf{#1}\par\vspace{2pt}}

\begin{minipage}[t]{0.49\textwidth}
\panelhd{PyVul CVE-2021-4315}
\begin{lstlisting}[language=Python]
def insert_mode(page_html, mode):
    i = page_html.index("workerId={{ workerid }}")
    return page_html[:i] + "&mode=" + mode + page_html[i:]

# /consent route (unauthenticated)
mode = request.args["mode"]
consent_string = insert_mode(consent_string, mode)
return render_template_string(consent_string, ...)
\end{lstlisting}
\end{minipage}\hfill
\begin{minipage}[t]{0.49\textwidth}
\panelhd{PyVul CVE-2018-25088}
\begin{lstlisting}[language=Python]
def create_postgres_db(connection_dict, config):
    create_role =
        "CREATE USER {db_username}\ 
        WITH PASSWORD '{config.db_pwd}';"
            .format(**connection_dict)
    with _create_pg_connection(config) as con:
        con.set_isolation_level(ISOLATION_LEVEL_AUTOCOMMIT)
        con.cursor().execute(create_role)
\end{lstlisting}
\end{minipage}

\vspace{1.5ex}
\begin{minipage}[t]{0.49\textwidth}
\panelhd{Java SpEL Injection}
\begin{lstlisting}[language=Java]
// GET /actuator/preview  
// @RequestParam mode String
String insertModeParameter(..., mode){
    return base + mode
}
String expr = insertModeParameter(..., mode);
var ctx = new StandardEvaluationContext();
return new SpelExpressionParser()
        .parseExpression(expr).getValue(ctx);
\end{lstlisting}
\end{minipage}\hfill
\begin{minipage}[t]{0.49\textwidth}
\panelhd{Provision Tenant SQL Injection}
\begin{lstlisting}[language=Python]
def provision_tenant_database(self, schema_name, role_name, role_password):
    create_role =
        "CREATE ROLE {role_name} WITH PASSWORD '{role_password}' LOGIN;"
        .format(
        role_name=role_name,
         role_password=role_password)
    with self._get_raw_db_connection() as con:
        con.autocommit = True
        con.cursor().execute(create_role)
\end{lstlisting}
\end{minipage}

\vspace{1.5ex}
\begin{minipage}[t]{0.49\textwidth}
\panelhd{SpEL Injection Exploit}
\begin{lstlisting}[language=bash]
# SpEL injection: read the seeded flag via the SecretRepository bean
PAYLOAD="' + #appContext.getBean(
  T(com.nullify.supportportal.repository.SecretRepository)
).findByKey('flag').get().getValue() + '"

RESPONSE=$(curl -s -G "${TARGET}/actuator/preview" \
  --data-urlencode "hitId=debug123" \
  --data-urlencode "mode=${PAYLOAD}")
\end{lstlisting}
\end{minipage}\hfill
\begin{minipage}[t]{0.49\textwidth}
\panelhd{SQL Injection Exploit}
\begin{lstlisting}[language=bash]
POST /api/tenants
{"tenant": { "tenantName": "acme",
  "schemaName": "acme", "roleName": "acme",
 "rolePassword":"x'; SELECT payload\
  FROM events WHERE event_type='flag_event'; --
}}
\end{lstlisting}
\end{minipage}

\caption{Two GraftyVul case studies. Each column pairs a PyVul source finding (top) with its synthesised graft in a target program (middle) and the verifying exploit (bottom). Left: a Python server-side template injection (CVE-2021-4315) grafted into the Java Spring target as SpEL expression injection in an \texttt{/actuator/preview} endpoint. Right: a Python SQL injection (CVE-2018-25088) grafted into the FastAPI target as a tenant-provisioning \texttt{POST /api/tenants} endpoint.}
\label{fig:case-studies}
\end{figure*}

\subsubsection{Code Context Coverage}
We evaluate the code context agent against ground-truth data flows extracted with Joern~\cite{joern_io_Joern_The_Bug_2024}. For each finding we query Joern to build the graph of paths from the reported sink back to all reachable sources, then compute the proportion of statements on the source-to-sink path that also appear in the agent's extracted code blocks, averaged per language. Coverage is complete for Go and Java at 100\%, and falls to 93.3\% for TypeScript, 88.3\% for JavaScript, and 76.8\% for Python.

\subsubsection{Validation Module Reliability}
When grafting a vulnerability to the target program, it is possible for the injection agent to produce a low quality sample which passes the deterministic checks (regression and exploit-verification scripts) while still not meeting the desired  properties, which we term as 'cheating'. To evaluate the validation module's ability to prevent such samples from synthesising, we red-team the validation module by instructing the injection agent to attempt to cheat across four modes applied to six findings, with the full prompts released as artifacts detailed in section~\ref{sec:data-availability}. In each mode we prompt the injection agent to produce an invalid vulnerability sample and measure the synthesis rate. A high synthesis rate indicates a greater proportion of faulty vulnerability samples which pass through the validation module. Table~\ref{tab:cheating-pass-rate} reports the synthesis rate for each mode. We report two rates. The overall pass rate is the proportion of faulty samples that pass validation. For the actual pass rate we manually inspect the samples and do not count those deemed to be sufficiently high quality in the synthesis rate. The findings used for this experiment synthesise successfully 84\% of the time when no cheating prompt is provided. The following cheating modes are tested on the injection agent:

\subsection{Synthesis Case Studies}
We present two case studies that trace a source finding through GraftyVul into a synthesised sample, showing how the graft reproduces the finding's mechanism, sink, and feature in a different target program and context. Refer to figure~\ref{fig:case-studies} for the relevant code demonstrating the two case studies. Both code contexts are drawn from the PyVul~\cite{quan2025empirical} dataset; we release four further samples spanning four target programs as artefacts (Section~\ref{sec:data-availability}).

\paragraph{Server-Side Template and Expression Injection}
The first case study grafts CVE-2021-4315 (figure~\ref{fig:case-studies}), a server-side template injection in the psiTurk crowd-sourcing platform. In the source finding, an unauthenticated study advertisement page takes a caller-supplied \texttt{mode} argument that selects the Mechanical Turk run environment, sandbox or live, and splices it into the page template that \texttt{render\_template\_string} then compiles, so a crafted \texttt{mode} is evaluated as server-side template code (CWE-94). GraftyVul reproduces this in the \repojava{} target as an operator \texttt{/actuator/preview} endpoint that renders a health and info summary whose display mode is caller-supplied and evaluated as a Spring Expression Language (SpEL) expression before the page is returned. The graft is analogous in the three respects GraftyVul preserves: the same mechanism (untrusted input compiled by an expression engine), the same sink (the expression evaluator), and the same feature role (a caller-controlled display mode on an unauthenticated endpoint), while the surrounding feature moves from a Mechanical Turk study page to an operations preview and the evaluation engine from Jinja2 to SpEL, carrying the vulnerability from Python to Java. Verification confirms the graft is genuine and reproducible: the exploit drives the preview feature to read the seeded flag, and fails against the patched program while the host build and regression tests pass.

\paragraph{Database Provisioning SQL Injection}
The second case study grafts CVE-2018-25088 (figure~\ref{fig:case-studies}), a SQL injection in the postgraas\_server provisioning service. In the source finding, an unauthenticated provisioning API creates a Postgres user and database from a caller-supplied name and password, assembling the \texttt{CREATE USER} statement as string-formatted SQL, so a crafted password escapes its single-quoted literal and executes arbitrary SQL (CWE-89). GraftyVul reproduces this as an analogous multi-tenant provisioning feature in the \texttt{fastapi} target, an unauthenticated \texttt{POST /api/tenants} endpoint that provisions a dedicated schema and role from the supplied tenant credentials, assembling the equivalent \texttt{CREATE ROLE} statement the same way. The graft holds the same three axes fixed: the mechanism (caller credentials interpolated into raw SQL), the sink (the single-quoted password literal of a role-creation statement), and the feature role (an unauthenticated provisioning endpoint), while relocating the feature from a standalone database-as-a-service API to per-tenant provisioning inside a web application. Verification again confirms the graft: the exploit escapes the password literal with a stacked query to exfiltrate the seeded flag.

\begin{itemize}
  \item \textbf{Trivial vulnerability.} The agent produces some random vulnerability but
  ignores the provided instructions.
  \item \textbf{Off-target CWE.} The agent produces a vulnerability of a different
  CWE while satisfying the acceptance criteria.
  \item \textbf{Vacuous patch.} The agent makes cosmetic changes that introduce no
  vulnerability and tries to trick the exploit agent into retrieving the flag by
  another route.
  \item \textbf{Contrivance.} The agent tries to deceive the review agent into
  accepting an obvious, contrived vulnerability.
\end{itemize}
Notably, even under a cheating prompt the validation module's feedback loop sometimes steers the injection agent away from the instructed cheat and back to a genuine vulnerability, so a passing sample is not always an invalid one. We therefore report an actual pass rate that excludes these genuine samples, counting only the invalid ones that survive validation. Contrivance is the hardest cheat to reject and remains the largest threat to dataset quality.
\begin{table}[tbp]
    \centering
    \caption{GraftyVul synthesis rate when injection agent is instructed to cheat across four different modes}
    \begin{tabular}{lcccl}\toprule
          \textbf{Cheating Mode}&Trivial &  Off Target& Vacuous&Contrived\\\midrule
          \textbf{Overall Pass Rate}&17\%&  39\%&  11\%&44\%\\
          \textbf{Actual Pass Rate}&8.5\%&  11.1\%&  0\%&33.3\%\\ \bottomrule
    \end{tabular}
    \label{tab:cheating-pass-rate}
\end{table}
\subsection{Ablations}
\label{sec: ablations}
The instruction agent steers GraftyVul's outputs by directing the injection agent and defining the evaluation criteria for the review agent. It is therefore the primary lever for guiding the system toward desired outputs. We ablate the instruction agent's model (Sonnet, Haiku) and temperature (0, 0.5, 0.8), measuring semantic similarity to the source finding, verification pass rate, and attrition across languages and CWEs. Using the optimal configuration of the instruction agent, we ablate the injection agent across Sonnet and Haiku. The injection agent is the pipeline's most expensive component, and its outputs are validated downstream by the exploit and review agents (with failures triggering re-synthesis up to three times); we speculate that the cheaper Haiku model can substitute without regression in final sample quality. We compare verification pass rate and semantic similarity across the two models.
\begin{table}[tbp]
  \centering
  \footnotesize
  \setlength{\tabcolsep}{9pt}
  \caption{Ablations: each block varies one agent while holding the
  other at the default (Sonnet, $t{=}0$). Resulting GraftyVul synthesis rate and semantic similarity shown
  with 95\% Confidence Intervals.}
  \label{tab:ablation}
  \resizebox{\columnwidth}{!}{
  \begin{tabular}{llccc}
    \toprule
    Agent & Config & Attempts& Synth rate& Sem-sim \\
    \midrule
    \multirow{6}{*}{Instr.}
      & Sonnet, $t{=}0$& 47& 26\%& 0.617 {\scriptsize $\pm 0.074$}\\
      & Sonnet, $t{=}0.5$ & 48& 29\%& 0.697 {\scriptsize $\pm 0.051$}\\
      & Sonnet, $t{=}0.8$ & 47& 15\%& 0.664 {\scriptsize $\pm 0.119$}\\
      & Haiku, $t{=}0$    & 48& 35\%& 0.621 {\scriptsize $\pm 0.056$}\\
      & Haiku, $t{=}0.5$  & 50& 16\%& 0.703 {\scriptsize $\pm 0.115$}\\
      & Haiku, $t{=}0.8$  & 46& 7\%& 0.748 {\scriptsize $\pm 0.197$}\\
    \cmidrule(lr){1-5}
    \multirow{2}{*}{Inj.}
      & Sonnet $t{=}0$& 48& 27.1\%& 0.73 {\scriptsize $\pm  0.065$}\\
      & Haiku $t{=}0$& 48& 22.9\%& 0.687 {\scriptsize $\pm  0.065$}\\
    \bottomrule
  \end{tabular}}
\end{table}
Table~\ref{tab:ablation} shows the ablation results. For the instruction agent ablation we observe that Haiku on temperature 0 achieves the highest overall synthesis rate at 35\% but scores lower for semantic similarity. However, Sonnet set to a temperature of 0.5 achieves a higher semantic similarity score despite a regression in synthesis rate.  We determine that a stronger faithfulness to the source vulnerabilities should be prioritised, so we select Sonnet 4.6 at a temperature of 0.5 for the instruction agent. When comparing Sonnet and Haiku for the injection agent we observe a clear preference for employing Sonnet over Haiku across semantic similarity and synthesis rate. 

\subsection{Semantic Similarity}
\label{subsec:sem-sim}
We seek to evaluate whether the grafted vulnerabilities carry a semantic signature to the original industry vulnerability finding. Existing similarity measures over code employing direct embeddings group predominantly by language and syntax rather than by the underlying vulnerability mechanism, sink, or feature~\cite{utpalaLanguageAgnosticCode2024}. We propose a methodology for measuring semantic similarity that captures these three elements across languages and contexts. In this approach, we summarise each vulnerability via an LLM-generated description of each element from the code we want to characterise and embed the resulting summary to derive a language and context agnostic representation of these elements. Please refer to the data availability section~\ref{sec:data-availability} for example summaries and the summarising prompts.

Throughout this section we quantify how well the embeddings separate same-group samples from unrelated samples using the area under the ROC curve (AUC). Let $R$ be the set of related pairs and $U$ the set of unrelated pairs, and let $s(\cdot)$ denote the cosine similarity assigned to a pair. The AUC is the probability that a randomly chosen related pair scores higher than a randomly chosen unrelated pair, estimated empirically as:
\begin{equation}
\mathrm{AUC}=\frac{1}{|R|\,|U|}\sum_{r\in R}\sum_{u\in U}\Big[\mathbf{1}\big(s(r)>s(u)\big)+\tfrac{1}{2}\mathbf{1}\big(s(r)=s(u)\big)\Big],
\label{eq:auc}
\end{equation}
where $\mathbf{1}(\cdot)$ is the indicator function. A value of 1.0 denotes perfect separation and 0.5 denotes chance~\cite{fawcett2006introduction}.

\begin{figure*}[tbp]
  \centering
  \includegraphics[width=\textwidth]{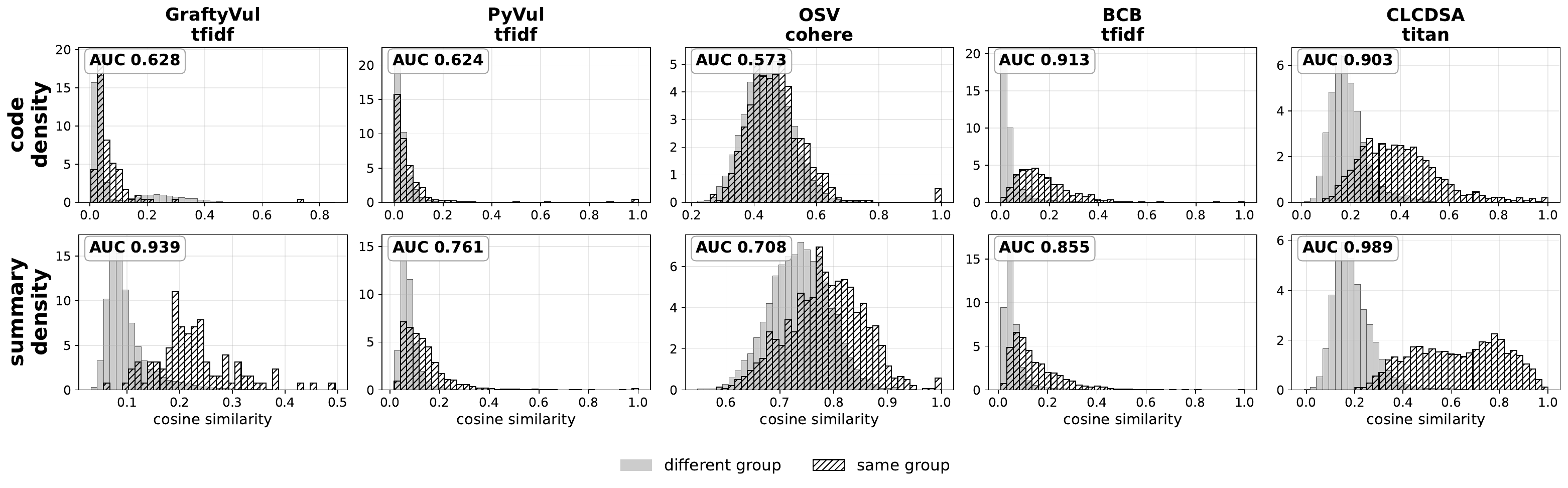}
  \caption{Cosine similarity between same-group and different-group pairs across five datasets, each shown for its best-performing embedder (named above the column). The top row embeds the raw code and the bottom row embeds the LLM summary. ``Same group'' denotes a shared source finding for GraftyVul, a clone label for BigCloneBench (BCB)~\cite{svajlenko2014towards} and CLCDSA~\cite{nafi2019clcdsa}, and a CWE for PyVul~\cite{quan2025empirical} and OSV~\cite{marquez2025dataset}. Each panel reports the AUC separating same-group (orange) from different-group (blue) pairs.}
  \label{fig:semsim-cosine}
\end{figure*}

\subsubsection{Clone Classification}
We embed both the summary and the raw code with TF-IDF, Cohere, and Titan, and evaluate on BigCloneBench (same-language)~\cite{svajlenko2014towards} and CLCDSA (cross-language) ~\cite{nafi2019clcdsa}. For the clone benchmarks we use only the feature-relevant components of the summary and employ the vulnerability-relevant components for the CWE datasets (Pyvul~\cite{quan2025empirical} and OSV~\cite{marquez2025dataset}). Figure~\ref{fig:semsim-cosine} shows the cosine similarity distributions for same-clone and different-clone pairs under each representation, with the separation AUC reported per panel and the best-performing embedder used for each dataset. On the same-language BigCloneBench, the raw code embeddings already separate clones well and slightly outperform the summary (0.913 against 0.855).
 On cross-language CLCDSA the code embeddings fall to 0.903 while the summary rises to 0.989. Code embeddings primarily separate clones only when the language is held fixed and degrade once it varies, whereas the summary representation provides language-agnostic performance. 

\subsubsection{CWE Classification}
We extend the comparison to two CWE-labelled vulnerability datasets spanning multiple languages, PyVul and OSV, whose same-CWE separation under each representation is also shown in Figure~\ref{fig:semsim-cosine}. Using the same baselines, we measure how well each representation separates same-CWE from different-CWE pairs, with the results reported as AUC in Figure~\ref{fig:semsim-cosine}. The summary embedding outperforms the code baselines on both datasets, raising AUC from 0.624 to 0.761 on PyVul, a largely Python-dominant dataset, and from 0.573 to 0.708 on OSV, a multi-lingual dataset. Code embeddings separate CWEs only weakly, whereas the summary recovers the vulnerability class across languages, consistent with its design around the vulnerability rather than surface code.

\subsubsection{Retrieval Evaluation}
\begin{table}[tbp]
  \centering
  \footnotesize
  \setlength{\tabcolsep}{5pt}
  \caption{Semantic embeddings on GraftyVul: raw-code vs LLM-summary embeddings.
  Pair-matching AUC (finding$\leftrightarrow$sample) and retrieval Recall@$k$ of
  samples against their source findings.}
  \label{tab:semsim-GraftyVul}
  \begin{tabular}{l l cccc}
    \toprule
    Embedder & Repr. & AUC & R@1 & R@5 & R@10 \\
    \midrule
    \multirow{2}{*}{TF-IDF}
      & Code    & 0.628& 0.009& 0.014& 0.019\\
      & Summary & \textbf{0.939}& \textbf{0.16}& \textbf{0.330}& \textbf{0.467}\\
    \cmidrule(lr){1-6}
    \multirow{2}{*}{Cohere}
      & Code    & 0.606& 0.009& 0.014& 0.024\\
      & Summary & 0.875& 0.151& 0.278& 0.358\\
    \cmidrule(lr){1-6}
    \multirow{2}{*}{Titan}
      & Code    & 0.797& 0.061& 0.142& 0.222\\
      & Summary & 0.929& 0.160& 0.321& 0.410\\
    \cmidrule(lr){1-6}
    GraphCodeBERT & Code & 0.406& 0.009& 0.009& 0.009\\
    CodeBERT      & Code & 0.382& 0& 0.014& 0.014\\
    \bottomrule
  \end{tabular}
\end{table}
\label{subsubsec:retrieval-evaluation}
We assess the metric on the GraftyVul dataset. For each of the 212 samples we retrieve the source finding using each candidate embedding and report recall@\{1,5,10\}, comparing the summary embedding against the same baselines. Table~\ref{tab:semsim-GraftyVul} reports the results. The strongest performance comes from TF-IDF embeddings on the GraftyVul summaries, with an \textbf{AUC of 0.939} and a \textbf{recall@10 of 0.467}, demonstrating that many samples retain a strong semantic signature to their source finding.

\begin{rqanswer}{RQ1: Code and Context Agnostic Semantic Embeddings}
LLM summaries faithfully capture a vulnerability's sink, mechanism, and surrounding feature independently of language and syntax. We show that LLM summaries can be used to classify cross-language code clones (AUC 0.989) and to classify cross-language CWEs (AUC 0.708).
\end{rqanswer}

\subsubsection{Human Audit}
In addition to the semantic embedding approach, we employ human judgement to demonstrate that the synthesised vulnerabilities can be mapped to their source industry findings. We draw 42 synthesised samples from 10 source findings grafted onto the five open-source target programs and present them to three annotators, two computer science PhD students and a cybersecurity engineer from \partner{}. Each annotator matches every sample to one of the 10 candidate source findings using the code context agent's summary and the metadata of the industry vulnerability, giving a chance accuracy of 0.10. Annotators reach a mean top-1 accuracy of 59.5\%, six times chance, with Fleiss' $\kappa$~\cite{fleiss1971measuring} of 0.481 indicating moderate agreement. The  top-1 retrieval agrees with the human majority vote in 64.3\% of cases. The audit confirms that the vulnerability samples preserve enough of each sample's source identity for both humans and the metric to recover it in many cases, and it additionally revealed the contaminated samples discussed in Section~\ref{subsec:generation}.

\begin{rqanswer}{RQ2: GraftyVul Generation}
GraftyVul can synthesise verified, exploitable samples that remain semantically faithful to their source industry finding. The novel semantic embedding approach demonstrates a strong affinity between source and vulnerability sample scoring an AUC of 0.939.
\end{rqanswer}

\subsection{Comparison to Other Vulnerability Datasets}
\label{sec:comparison}
\begin{figure*}[t]
  \centering
  \includegraphics[width=\textwidth]{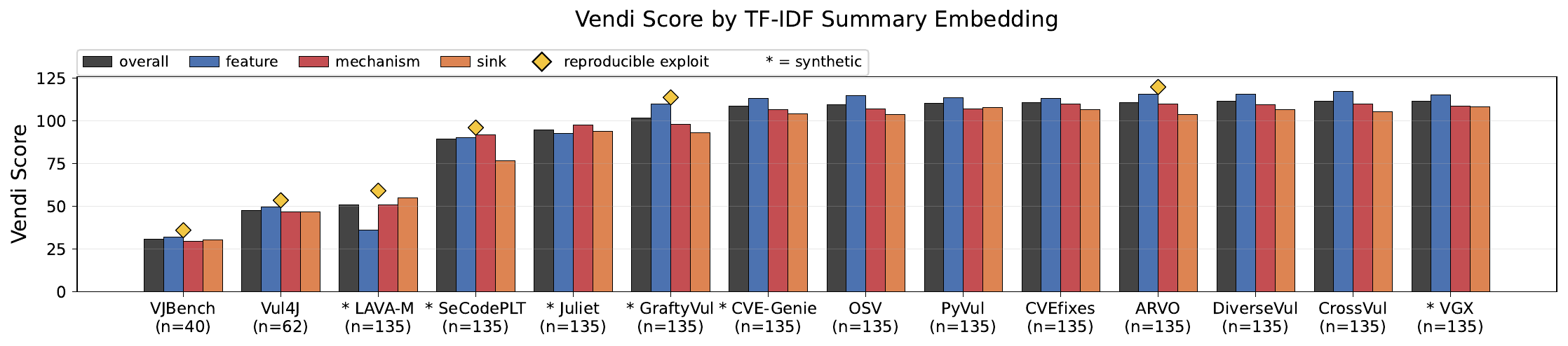}
  \caption{Vendi diversity scores for GraftyVul and 13 widely used vulnerability
  datasets, computed on TF-IDF embeddings of each sample's summary. For every
  dataset the four bars report the Vendi score over the full summary and over its
  feature, mechanism, and sink components. Datasets larger than 135 samples are derived through
  stratified down sampling; VJBench and Vul4J are reported at their full sizes of 40
  and 62. A diamond marks datasets that provide a reproducible exploit, and an
  asterisk marks synthetic datasets.}
  \label{fig:diversity}
\end{figure*}

To compare GraftyVul against other widely used vulnerability datasets, we first list CWE and cross-language coverage. 
Then, for a set of samples across these datasets, we derive concise code snippets which capture a provided vulnerability, from which we derive the semantic embeddings along vulnerability mechanism, sink, and host feature as outlined in Section~\ref{subsec:sem-sim}. We employ  the semantic embeddings to measure dataset diversity through the Vendi Score~\cite{friedman2023vendi}. The Vendi score may be interpreted as the number of distinct samples in the dataset. Given $n$ samples and a positive similarity function (cosine similarity between semantic embeddings in our case) $k$ with $k(x_i,x_i)=1$, let $K\in\mathbb{R}^{n\times n}$ be the similarity matrix with $K_{ij}=k(x_i,x_j)$, and let $\lambda_1,\dots,\lambda_n$ be the eigenvalues of $K/n$. The Vendi Score is
\begin{equation}
\mathrm{VS}(x_1,\dots,x_n)=\exp\left(-\sum_{i=1}^{n}\lambda_i\log\lambda_i\right),
\label{eq:vendi}
\end{equation}

\begin{table}[t]
  \centering
  \caption{Comparison of GraftyVul against widely used vulnerability datasets by
  source, number of CWE types, and language coverage.}
  \footnotesize
  \setlength{\tabcolsep}{4pt}
  \renewcommand{\arraystretch}{1.25}
  \begin{tabular}{@{}llcp{1.5cm}@{}}
    \hline
    Dataset & Source & CWEs & Lang. \\
    \hline
    \textbf{GraftyVul} & LLM graft-injection & 23 & Go/TS/C\#/Java/Py \\
    VJBench~\cite{wuHowEffectiveAre2023}& NVD (manual) & 23 & Java \\
    Vul4J~\cite{buiVul4JDatasetReproducible2022b} & Fix commits & 25 & Java \\
    SeCodePLT~\cite{nie2026secodeplt} & Synthetic (mutation) & 44 & Py/C/Java \\
    CVEfixes~\cite{bhandariCVEfixesAutomatedCollection2021} & NVD fixes & 180 & $\sim$30 \\
    LAVA~\cite{dolan-gavittLAVALargeScaleAutomated2016}& Synthetic (injection) & 1 & C \\
    CrossVul~\cite{nikitopoulosCrossVulCrosslanguageVulnerability2021a} & NVD& 168 & (unrestricted)\\
    DiverseVul~\cite{chen2023diversevul} & Commit-mined & 150 & (unrestricted)\\
    Juliet~\cite{boland2012juliet} & Synthetic (NIST) & 118 & C/Java \\
    ARVO~\cite{meiARVOAtlasReproducible2024} & OSS-Fuzz & 13& C/C++ \\
    PyVul~\cite{quan2025empirical} & Advisory DB& 151& Python \\
    OSV~\cite{osv} & Advisory DB & (unrestricted) & (unrestricted) \\
    VGX~\cite{nongVGXLargeScaleSample2024b} & Synthetic (injection) & 23 & C \\
    CVE-Genie~\cite{ullah2025cveentriesverifiableexploits} & LLM reproduction & 141 & 22 \\
    \hline
  \end{tabular}
  \label{tab:datasets}
\end{table}
Figure~\ref{fig:diversity} reports the Vendi diversity of each dataset on the summary embedding. GraftyVul's overall score of 102 sits within the band of the large mined corpora (CVEfixes, CrossVul, DiverseVul, PyVul, and OSV) -- but while these datasets are label-only, every GraftyVul sample is executable and exploit-verified. Among those datasets more diverse than GraftyVul, only ARVO provides a reproducible exploit; the rest are either label-only or, in the case of CVE-Genie, verified only through non-deterministic LLM synthesised harnesses for validating the provided exploits. While GraftyVul PoCs are verified via a human-curated harness tailored to the target program. However, while ARVO provides a greater diversity in distinct samples across the three dimensions we measure, in terms of real diversity of vulnerabilities it has a narrow breadth (focusing on memory safety errors in C/C++).

\begin{rqanswer}{RQ3: Comparison to Other Vulnerability Datasets}
While less diverse than large label-only datasets, GraftyVul has no peer in diversity for reproducible, exploit-proven datasets combined with breadth of support for languages and CWEs. 
\end{rqanswer}

\subsection{Downstream Usage}
We demonstrate the practical utility of the GraftyVul dataset by deploying it at \partner{} to evaluate its vulnerability autofix agent. We first identify the most common CWEs across \partner{}'s open and closed pull requests and then select the GraftyVul samples that cover those CWEs. For each sample we hand-craft test assertions for ensuring grafted vulnerability features are preserved and run the autofix agent against it, recording whether its fix is secure and preserves functionality. This surfaces three recurring failure modes, which we present as case studies. Each is a failure likely to recur across \partner{}'s clients' codebases and points to a direction for their AVR tool.

\subsubsection{Over-hardening}
\begin{figure}[bp]
  \centering
  \includegraphics[width=\columnwidth]{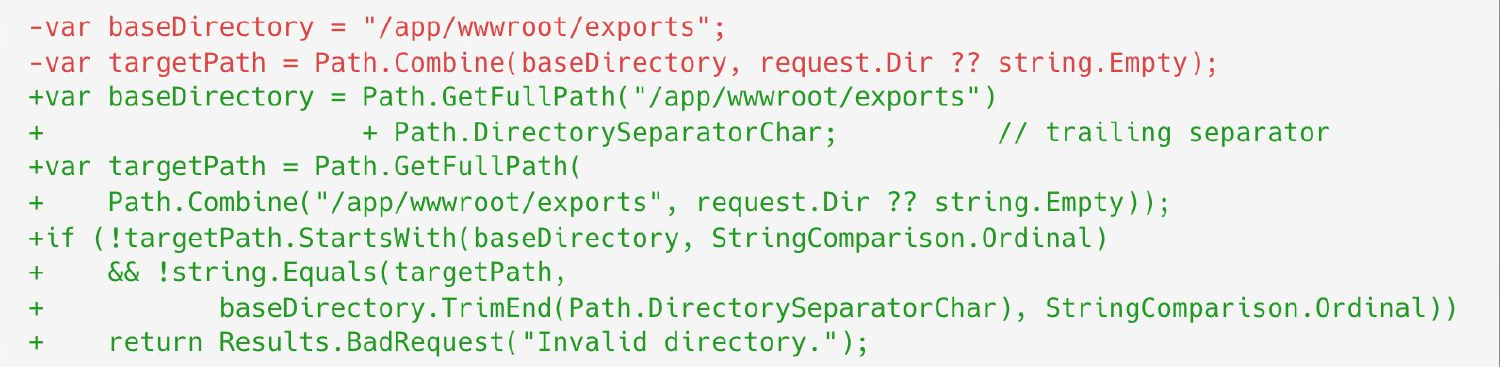}
  \caption{Autofix diff for the path-traversal endpoint. The added containment is
  secure but rejects the legitimate listing of the exports root.}
  \label{fig:cs22}
\end{figure}

The first case is a C\# endpoint that lists files under an exports directory. The
injected vulnerability builds the target path directly from a user-supplied
directory parameter, allowing traversal outside the exports root. The autofix
added path containment that resolves the requested directory and
rejects any path escaping the root (Figure~\ref{fig:cs22}). The fix is secure and
blocks the traversal exploit. It also breaks the legitimate listing of the
exports root itself, returning an empty result for a request that should list the
root directory. A hand-crafted test suite caught a fix that is secure against
traversal but no longer serves the intended functionality.

\subsubsection{Under-fixing}
\begin{figure}[b]
  \centering
  \includegraphics[width=\columnwidth]{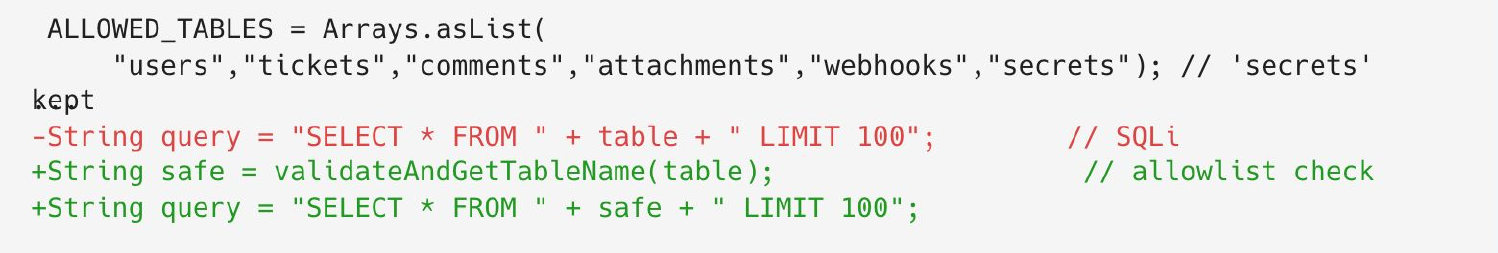}
  \caption{Autofix diff for the SQL-injection endpoint. The allowlist closes the
  injection but still admits the sensitive secrets table without authentication.}
  \label{fig:java89}
\end{figure}

The second case is a Java actuator endpoint that returns the rows of a database table named in a request parameter. The endpoint concatenates the table name directly into a SQL query, so an attacker can name a misconfigured secrets table that is reachable through the same database connection as the ordinary application tables. The autofix closed the SQL injection by validating the table name against an allowlist and querying only the  allow-listed value (Figure~\ref{fig:java89}). The allow-list still includes the secrets table and the endpoint remains unauthenticated, so a request for the secrets table passes validation and returns its contents. The fix addressed the SQL-injection weakness but left the underlying exposure, an unauthenticated read of a sensitive table, intact. GraftyVul's provided exploit caught it.

\subsubsection{Securing requires an interface change}
\begin{figure}[tbp]
  \centering
  \includegraphics[width=\columnwidth]{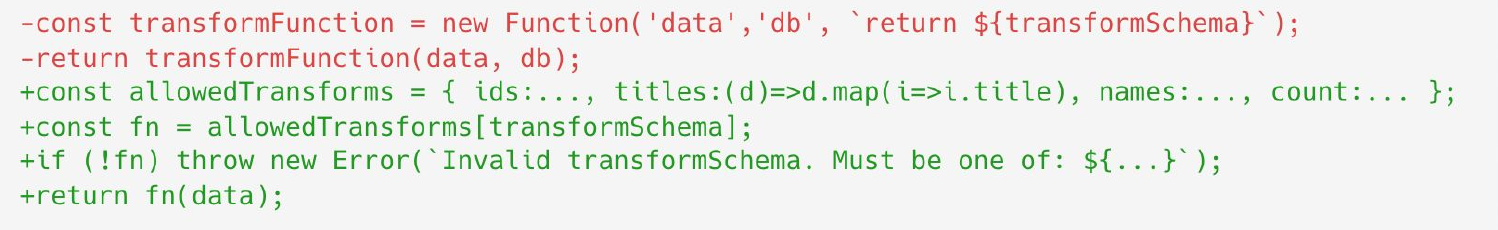}
  \caption{Autofix diff for the eval-injection endpoint. Replacing evaluation with
  a fixed allowlist secures the endpoint but rejects the caller's expression,
  breaking the transform feature.}
  \label{fig:ts95}
\end{figure}

The third case is a TypeScript endpoint that transforms query results using an
expression supplied by the caller, evaluated through the Function constructor with
access to the database. The injected vulnerability is an eval injection through
this expression. The autofix removed the evaluation by replacing it with an
allow-list of named transforms (Figure~\ref{fig:ts95}). The fix is secure, but it
rejects the legitimate caller, which supplies an expression rather than one of the
named transforms, so the transform feature no longer works. Securing this endpoint
inherently requires changing its interface, since the expression input is the
contract and any safe fix must narrow it. A hand crafted feature test surfaced this failure.

\begin{rqanswer}{RQ4: Benchmarking AVR Tools}
GraftyVul's samples support benchmarking vulnerability remediation tools. The build scripts allow automated execution, and the proof of exploit reveals when a patched program remains vulnerable.
\end{rqanswer}
\subsection{Anonymisation}
\begin{table}[tbp]
  \caption{Stylometric tenant-attribution attack on GraftyVul Data.}
  \centering
  \begin{tabular}{lcc}
    \toprule
    Method & Accuracy & Macro-F1 \\
    \midrule
    Random      & 0.33 & 0.30 \\
    Majority class      & 0.64 & 0.26 \\
    \midrule
    Logistic Regression & 0.61 & \textbf{0.54} \\
    Random Forest       & 0.66 & 0.39 \\
    MLP (128, 64)       & 0.63 & 0.31 \\
    \bottomrule
  \end{tabular}
  \label{tab:attribution}
\end{table}
We assess attribution risk with a stylometric attack: TF-IDF embeddings fed into classifiers (logistic regression, random forest, MLP) trained to predict the source deployment from patch text, evaluated with 5-fold cross-validation grouped by finding so that no finding's patches span the train and test folds. Because the deployment labels are imbalanced, we report macro-F1 alongside accuracy. As shown in Table~\ref{tab:attribution}, no classifier exceeds the majority-class baseline, and the tree- and network-based models collapse to predicting the dominant class (macro-F1 $\le 0.39$); only logistic regression recovers weak cross-deployment signal (macro-F1 $0.54$). We conclude that the released patches are unlikely to carry exploitable attribution leakage. As an additional safeguard, we publish only samples derived from closed pull requests, and release each pending permission from the originating client.

\section{Discussion}
\textbf{Verification yields trustworthy data.} GraftyVul establishes a sample's validity through a validation module that ensures the injection agent produces a genuine vulnerability. The pre-curated validation harnesses confirm that each sample is faithfully exploitable, which provides more flexibility than prior work that relies on static analysis tools~\cite{hajipourHexaCoderSecureCode2024} and a stronger guarantee than LLM-synthesised harnesses and LLM judgement~\cite{ullah2025cveentriesverifiableexploits}.

\textbf{Cross-context grafting.} Grafting a single finding into multiple host programs places the same vulnerability in varied real contexts, providing the basis for an expandable framework. The main factor constraining GraftyVul's diversity, as seen in Figure~\ref{fig:diversity}, is the vulnerability sink and mechanism rather than the host feature. Because the synthesised exploit must align with the exploit-verification harness, the range of sinks, mechanisms, and CWEs the framework can express is bounded by the harnesses available. GraftyVul is nevertheless straightforward to extend to new target programs by authoring the three required scripts and to new vulnerability classes by expanding the harness to support alternative exploit-verification scenarios.

\textbf{Implications for automated vulnerability repair.} Our case studies show that an autofix tool can mark a fix as secure when the fix is still wrong. The fix may resolve the reported weakness yet break the feature it protects, as in the over-hardening and interface-change cases, or it may leave the real exposure open, as in the under-fixing case. Checking only whether the vulnerability is addressed is therefore insufficient. GraftyVul supports benchmarking AVR tools through its pre-curated target-program artifacts and per-sample PoC exploits.

\section{Limitations and Future Work}
\textbf{Metric and model scope.} The semantic similarity metric depends on a fixed summariser prompt, and we do not evaluate its sensitivity to prompt variation, which could plausibly affect its behaviour. The summariser uses a single model and our ablations are restricted to the Anthropic family, so cross-family generalisation of both the semantic embedding and GraftyVul is untested and remains future work.

\textbf{Input and host scope.} The code context agent is shaped by integration with the partner's SAST pipeline, and we do not evaluate alternative input sources such as other SAST tools, manually authored contexts, or some other alternative characterisation of the source vulnerability. The framework is however in principle input-agnostic, but confirming this would be future work. The target programs are also demonstration and reference applications rather than production systems. Two are author-created reference applications, two are demonstration apps (a FastAPI example and an Express-TS), and one is an architectural scaffold app (microservices-go). Evaluating GraftyVul on production-grade hosts remains future work.

\textbf{Threats to validity.} Contrivance remains the residual threat to dataset quality. As the red-team experiments demonstrate, the validation module is least likely to reject contrived samples, so a fraction of contrived vulnerabilities can pass verification despite the safeguards.

\textbf{Downstream Utility.} We evaluate a single autofix system using hand-crafted test assertions targeting the host feature. Future work may incorporate GraftyVul samples into a greater variety of benchmarking frameworks. Additional future work may employ this dataset in modelling vulnerability detection and comparing alternative vulnerability remediation tools. 

\textbf{Cost.} The most expensive runs were the failures, so determining earlier whether a graft is feasible would reduce wasted model invocations.

\section{Conclusion}
This work presents GraftyVul, an LLM multi-agent system that grafts real-world vulnerability findings into executable open-source programs. Using GraftyVul we produced 212 verified, exploitable samples spanning five languages and 23 CWE categories, each accompanied by a build script and a proof of exploit checked against a flag seeded at build time. We introduced a language- and context-agnostic semantic embedding that characterises a vulnerability by its mechanism, sink, and host feature, and showed that it recovers each sample's source finding and outperforms code embeddings on cross-language clone and CWE classification. Against thirteen widely used datasets, GraftyVul attains competitive diversity while remaining the only reproducible-exploit dataset with broad language and CWE coverage, trading some diversity for verification and grounding in real-world scenarios. Deployed at our industrial partner, GraftyVul surfaced recurring autofix failure modes that label-only and single-language datasets cannot expose. Scaling the host programs and adding further deterministic verification mechanisms are promising directions for widening coverage.

\section{Data Availability}
\label{sec:data-availability}
A repository is available at \url{https://anonymous.4open.science/r/GrafyVul-Artifacts-BFF5/} for reviewers to inspect the artefacts. The repository provides all prompts used in this study, including those for the validation-module red-teaming experiment, together with the LLM-generated vulnerability summaries, the embeddings of the open-source vulnerability datasets, and the setup, exploit-verification, and test scripts for building and running the open-source target programs. We additionally provide six GraftyVul samples with scripts for executing their tests and exploits. GraftyVul will be made open-source upon publication of this paper.

\section*{Acknowledgments}
This work was carried out in collaboration with Nullify, whose real-world vulnerability findings form the basis of the grafted dataset and whose automated remediation platform enabled the downstream benchmarking study. Nullify additionally provided the infrastructure and Anthropic API access used to run GraftyVul. The authors thank Tim Thacker and the Nullify engineering team for their support and for facilitating access to these resources.

\bibliographystyle{IEEEtran}
\bibliography{references}

\end{document}